\documentclass[12pt]{article}
\usepackage{setspace}
\usepackage{xurl,amsmath,amssymb,graphicx}
\usepackage{longtable,tabularx}
\begin{document}
\author{J. I. Katz\\Department of Physics and\\ McDonnell Center for the
Space Sciences\\ Washington University, St. Louis, Mo. 63130 USA\\
Orcid: 0000-0002-6537-6514 {\tt katz@wuphys.wustl.edu}}
\title{Painting Asteroids for Planetary Defense}
\maketitle
\begin{abstract}
	Asteroidal impact threats to the Earth will be predicted a century
	or more in advance.  Changing an asteroid's albedo changes the
	force of Solar radiation on it, and hence its orbit.  Albedo may be
	changed by applying a thin ($\sim 0.1\,\mu$) reflective coat of
	alkali metal, dispensed as vapor by an orbiting spacecraft.  A
	complete coat reduces the effective Solar gravity, changing the
	orbital period.  A Tunguska-class (50 m diameter) asteroid in a
	nominal orbit with perihelion 1 AU and aphelion 3 AU ($a = 2\,$AU,
	$e = 0.5$) may be displaced along its path by $\sim 1000\,$km in 100
	years, sufficient to avoid impact in a populated area, by
	application of one kg of lithium or sodium metal over its entire
	surface.  Alternatively, coating one hemisphere of an asteroid in an
	elliptical orbit may produce a Solar radiation torque, analogous to
	but distinct from the Yarkovsky effect, displacing it by an Earth
	radius in $\sim 200$ years.  The time required scales as the square
	root of the asteroid's diameter (the 1/6 power of its mass) because
	the displacement increases quadratically with time, making it
	possible to prevent the catastrophic impact of a km-sized asteroid
	with a minimal mass.
\end{abstract}
Keywords: planetary defense, asteroids
\newpage
\section{Introduction}
The Tunguska event, believed to be produced by impact of a rocky asteroid
with diameter 50--60 m, had an airburst yield estimated (uncertainly) as
$\sim 10$ megatons high-explosive equivalent \cite{T21} and blew down about
$10^3$ km$^2$ of forest.  Such events may occur every few hundred to a
thousand years \cite{S83,B02}, and would be very destructive were they to
occur over a city.

Extensive surveys \cite{B21} are being conducted to find and catalogue
asteroids (Near Earth Objects; NEO) whose orbits bring them \cite{A21,M21}
close enough to the Earth to present a threat of collision.  In the
forseeable future, such surveys are expected \cite{H21} to be complete to
sizes smaller than the Tunguska impactor, and the progress of observational
technology will extend them to yet smaller asteroids.

Once a NEO is found and its orbital elements determined, it will be possible
to predict whether it will or not collide with the Earth for centuries
because NEO orbits are very close to those of Newtonian point masses.
Orbital determination and monitoring of orbital modification with extreme
accuracy are possible if a transponder or transmitter is emplaced on or
orbits the target asteroid.

Preventing such impacts is the problem of ``planetary defense''.  An
asteroidal orbit may be changed by imparting momentum.  Methods proposed
include landing small ion thrusters or impact of a spacecraft.  Because
warning times are long, even tiny changes in the asteroidal velocity may be
sufficient to prevent collision, provided the orbit is changed long before
the predicted collision.  If there is no early warning (warning of cometary
impacts is likely to occur only when the comet has a heliocentric distance
$\lessapprox 5\,$AU, about a year before impact) ablation by a nearby
nuclear explosion may be necessary.

This paper investigates the possibility of changing an asteroidal orbit by
coating it with a reflective metal to change the force of Solar radiation
pressure on it.  This is analogous to the Yarkovsky effect, but differs in
that it depends on the instantaneous force of radiation pressure rather than
the delayed re-emission of absorbed energy.  The radiation pressure force
has the same inverse square dependence on distance as gravity, so the effect
of changing the albedo over the entire asteroidal surface would be
dynamically equivalent to a small change in the Solar mass.  Alternatively,
changing the albedo of one hemisphere of an asteroid in eccentric orbit may
produce a systematic trend in its angular momentum and a displacement
proportional to the square of the elapsed time.

It may not be necessary to avoid collision entirely; deflecting an asteroid
of the size that produced the Tunguska event to the ocean or to an impact
point hundreds of km from any population may be sufficient.  Because
asteroidal orbits are accurately calculated centuries in advance, the area
around the impact point could be used as farmland or nature reserve with
minimal infrastructure, and completely evacuated when impact is imminent.

A related proposal \cite{H10a,H10b} has suggested changing the albedo of
99942 Apophis to modify its Yarkovsky effect and deflect its trajectory.
Although this shares with the present work its dependence on radiation
pressure forces, the orbital mechanics differ.  Apophis and 65803 Didymos
\cite{D21a,D21b} are easier targets to reach and their orbits can be
influenced more rapidly than those of most asteroids because of their
smaller semimajor axes, but the principles are the same.
\section{Angular Momentum Conserving Forces}
\label{conserving}
\subsection{Reducing the Effective Solar Mass}
The magnitude of the fractional change in effective Solar mass is given by
the ratio of the radiation pressure force to the gravitational force:
\begin{equation}
	\label{epsilon}
	\begin{split}
	\epsilon &= {f (L_\odot/4 \pi R^2 c) \pi D^2/4 \over G M_\odot \pi
	D^3\rho_{ast}/(6R^2)} = {3 f \over 8}{L_\odot \over G M_\odot c D
	\rho_{ast}}\\ &\approx 1.0 \times 10^{-8} {f \over 0.4}{\text{50 m}
	\over D}{\text{3 g/cm}^3 \over \rho_{ast}},
	\end{split}
\end{equation}
where $L_\odot$ is the Solar luminosity, $M_\odot$ the Solar mass, $f$ the
difference in momentum albedo (that allows for the angular distribution of
the scattered light) between the bare asteroid and its reflective coat,
averaged over its surface, $R$ the distance from the Sun, $D$ the diameter
of the asteroid, $\rho_{ast}$ its density, $G$ the constant of gravity and
$c$ the speed of light.

Asteroids are generally dark, with energy albedos in the range 0.03--0.2,
while alkali metals are reflective, with energy albedos in the range
0.90--0.95 \cite{N16}.  Their difference $\Delta A \approx 0.8$.  However,
the momentum albedo of a spherical surface depends on the angular
distribution of its scattered light.  The momentum albedo of a specularly
reflecting sphere is identical to that of an absorbing sphere, so that
coating with a reflective metal would not change the radiation pressure
force; if the bare asteroid were perfectly absorbing and the coated surface
a specular reflector $f = 0$.  A rough but reflective asteroidal surface
might scatter according to Lambert's law (isotropically), leading to $f = 2
\Delta A/3 \approx 0.6$; in Eq.~\ref{epsilon} $f$ has been scaled to an
intermediate value.

The fact that changing the reflectivity of an asteroid changes the Solar
radiation force on it offers the opportunity to change its path by applying
a very thin coat of reflective material.  The effective mass of the Sun
would then be
\begin{equation}
	M^\prime = (1 - \epsilon)M_\odot.
\end{equation}
\subsection{Orbital Mechanics}
A complete treatment is complex, so we consider only simple limiting cases.
\subsubsection{Adiabatic Change of Reflectivity}
The asteroid moves in a central force field, so its angular momentum $\ell$
is conserved.  If the change in effective mass occurs adiabatically (over
many orbits) then the orbital eccentricity
\begin{equation}
	\label{e}
	e = \sqrt{1 + {2 E \ell^2 \over G^2 M_\odot^2}},
\end{equation}
where $E$ is the orbital energy per unit mass of the asteroid, would be
conserved.  From Eq.~\ref{e},
\begin{equation}
	{E^\prime \over E} = \left({M^\prime \over M_\odot}\right)^2 \approx
	1 - 2\epsilon.
\end{equation}
From the general relation
\begin{equation}
	a = -{GM_\odot \over 2 E}
\end{equation}
we find
\begin{equation}
	{a^\prime \over a} = {M^\prime \over M_\odot}{E \over E^\prime}
	\approx 1 + \epsilon.
\end{equation}
The orbital period
\begin{equation}
	P = 2 \pi \sqrt{a^3 \over GM_\odot}
\end{equation}
becomes $P^\prime$:
\begin{equation}
	{P^\prime \over P} = \left({a^\prime \over a}\right)^{3/2}
	\left({M^\prime \over M_\odot}\right)^{-1/2} \approx 1 + 2\epsilon.
\end{equation}
\subsubsection{Instantaneous Change of Reflectivity}
An impulsive (on a time scale $\ll P$) change in reflectivity changes both
$a$ and $e$, but $\ell$ is still conserved because the force is central.
If this occurs at perihelion 
\begin{equation}
	\ell = vr = va(1-e),
\end{equation}
where $v$ is the asteroid's velocity, perpendicular to the radius vector at
perihelion.  This velocity may be found from the condition that changing
its reflectivity does not change its energy:
\begin{equation}
	E = {v^2 \over 2} - {GM_\odot \over a(1-e)} = -{GM_\odot \over 2a}
\end{equation}
or
\begin{equation}
	\label{v}
	v^2 = {GM_\odot \over a}{1 + e \over 1 - e}.
\end{equation}

Following the instantaneous reduction in effective Solar mass to $M^\prime$,
the new energy (per unit mass of the asteroid)
\begin{equation}
	\begin{split}
		E^\prime &= {v^2 \over 2}-{G M^\prime \over a(1-e)} =
		{GM_\odot \over 2a} \left({1+e \over 1-e} - 2{M^\prime \over
		M_\odot(1-e)}\right)\\ &= {GM_\odot \over 2a}
		\left({1+e-2(1-\epsilon) \over 1-e}\right)
		= {GM_\odot \over 2a} \left(-1 + {2 \epsilon \over 1-e}
		\right)\\ &= E\left(1 - {2 \epsilon \over 1-e}\right).
	\end{split}
\end{equation}
Then
\begin{equation}
	a^\prime = -{G M^\prime \over 2 E^\prime} = a{(1-e)(1-\epsilon)
	\over 1-e-2\epsilon} = a\left(1 + \epsilon{e+1 \over 1-e}\right).
\end{equation}
Because the change in effective mass does not affect the perihelion (the
velocity remains perpendicular to the radius vector)
\begin{equation}
	a(1-e) = a^\prime(1-e^\prime)
\end{equation}
and
\begin{equation}
	e^\prime = e\left(1 + \epsilon{e+1 \over e}\right).
\end{equation}

The ratio of new to former period
\begin{equation}
	\begin{split}
	{P^\prime \over P} &= \left({a^\prime \over a}\right)^{3/2}
	\left({M^\prime \over M_\odot}\right)^{-1/2} = \left({(1-\epsilon)
	(1-e) \over 1-e-2\epsilon}\right)^{3/2} (1-\epsilon)^{-1/2}\\
	&\approx 1+{2+e \over 1-e}\epsilon.
	\end{split}
\end{equation}
If the change in albedo occurs instantaneously at aphelion $e$ is replaced
by $-e$ and the period ratio
\begin{equation}
	{P^\prime \over P} \approx 1 + {2-e \over 1+e}\epsilon.
\end{equation}
\subsection{Avoiding Collision}
The preceding results may be summarized:
\begin{equation}
	{P^\prime \over P} = 1 + \alpha
\end{equation}
where
\begin{equation}
	\label{alpha}
	\alpha \approx
	\begin{cases}
		{2+e \over 1-e}\epsilon &= 5 \epsilon\quad\text{perihelion}\\
		{2-e \over 1+e}\epsilon &= 1 \epsilon\quad\text{aphelion}\\
		&\phantom{=}\phantom{0} 2 \epsilon\quad\text{adiabatic},\\
	\end{cases}
\end{equation}
and the numerical factors were evaluated for $e = 0.5$.

Multiplying the orbital period by a factor $1 + \alpha$ changes the mean
anomaly $n$ after $N$ orbits by $\Delta n \approx 2 \pi \alpha N$ radians
and displaces the asteriod after a time $T$ by a distance
\begin{equation}
	\label{DX}
	\Delta X \sim a \Delta n \approx 2 \pi a \alpha N \approx 2 \pi a
	\alpha {T \over P},
\end{equation}
that necessarily varies (Eq.~\ref{DX} is not an equality unless $e=0$)
along its orbit.  Because the paths of Earth and asteroid intersect
obliquely, delaying the asteroid ($P^\prime > P$) moves the impact point on
the Earth's surface by $\sim \Delta X$.  To avert collision entirely would
require
\begin{equation}
	\Delta X \gtrsim R_E,
\end{equation}
where $R_E$ is the radius of the Earth (for a typical asteroidal relative
velocity $\sim 20\,$km/s gravitational focusing by the Earth is not large);
this relation would be an equality if the undeflected asteroidal trajectory
were through the center of the Earth.

This leads to a condition on $\alpha$ (and on $\epsilon$ from
Eq.~\ref{alpha})
\begin{equation}
	\alpha > {\Delta X P \over 2 \pi a T} \approx 2 \times 10^{-8}
	\left({\Delta X \over \text{1000 km}}\right)
	\left({\text{100 y} \over T}\right)\left({P \over \text{3 y}}\right)
	\left({\text{1 AU} \over R}\right).
\end{equation}
Comparison to Eqs.~\ref{epsilon} and \ref{alpha} indicates that a 50 m
diameter asteroid may be deflected by $\sim 3000\,$km in a century or 1000
km in $\sim 30\,$years.  Tunguska-class asteroids may be deflected enough
to miss cities and directed into a depopulated area, perhaps an ocean.
Because there would be a century of warning, that area could be used for any
purpose, such as agriculture or a nature reserve, that did would not require
major permanent infrastructure, and evacuated only when impact is imminent.
If the initial trajectory were through the center of the Earth it would
require 200--300 years to avoid collision entirely.
\section{Angular Momentum Non-Conserving Forces}
If the perturbation conserves angular momentum (Sec.~\ref{conserving}) the
displacement of the asteroid is proportional to the time elapsed
(Eq.~\ref{DX}).  However, if the perturbation produces a non-zero mean
torque the displacement would be quadratic in the time elapsed.  It may
exceed the Earth's radius in the characteristic warning times of a century
or two, enabling impact to be avoided entirely, and also allow for
uncertainty in the predicted trajectory.

For simplicity, consider the rotational angular momentum to lie in the
orbital plane.  Suppose reflecting material be applied to a hemisphere
containing one rotational pole of the asteroid, but not to the other
hemisphere, and that the rotational angular momentum have a component in
the asteroid's orbital plane.  If the rotational period is short compared to
the orbital period the reflectivity may be averaged over longitude so that
it depends only on latitude.  In a circular orbit the radiation torque
increases the angular momentum during half of the orbit and decreases it
during the other half, the two effects cancelling.

However, asteroidal orbits are significantly eccentric; a NEO with
perihelion of 1 AU and aphelion of 3 AU, perhaps representative, has an
eccentricity $e = 0.5$.  The ratio of the radiation pressure torque at
aphelion to that at perihelion is $(1-e)/(1+e) = 1/3$ while the ratio of
times spent near these points is $[(1-e)/(1+e)]^{-3/2} = 3^{3/2}$.
If the rotation axis were parallel to the semimajor
axis of the orbit no torque would be exerted around perihelion or aphelion
and angular momenta imparted between aphelion and perihelion would cancel
that imparted between perihelion and aphelion, so no cumulative angular
momentum would be imparted.  In contrast, if the rotational angular velocity
has a component parallel to the semiminor axis of the orbit the angular
momentum imparted by radiation pressure around aphelion may be nearly twice
that imparted (in the opposite direction) around perihelion; there would be
a net change of orbital angular momentum each orbit.

In the latter case, any anti-symmetric (about the rotational equator),
longitudinally averaged, distribution of reflectivity would produce a
cumulative change of orbital angular momentum, semi-major axis and period,
and a displacement increasing quadratically with time.  Such an effect may
occur naturally, and should be included in the orbital solutions of small
bodies.  It is distinct from the classical Yarkovsky effect because it
occurs only in elliptic orbits, requires a non-zero component of rotational
angular momentum in the orbital plane, and does not depend on a delay
between absorption and reradiation of energy.  Coating one hemisphere around
the axis of rotation with reflective material can change the orbital angular
momentum an order of magnitude more rapidly than the change resulting from
any natural asymmetry of the reflectivity; that natural asymmetry would be
included in the calculated ephemeris that predicts impact while the change
could displace the asteroid from an impact trajectory to a close approach
without collision.

The orbital parameters of the threat object (if search is extended to
sufficiently small objects an impactor will be found \cite{S83,B02}) are
not known prior to its discovery, so only a parametrized estimate can be
made.  The change in angular momentum in one orbit may be described by an
equivalent impulsive change in velocity at perihelion
\begin{equation}
	\begin{split}
		\Delta v &= g {\pi \over 4}D^2 {L_\odot \over 4 \pi R^2 c} 
	{\sqrt{R^3/GM_\odot} \over \pi \rho_{ast}D^3/6} = {3 g \over 8 \pi}
		{L_\odot \over \rho_{ast} D c \sqrt{GM_\odot R}}\\ &=
		0.023 g {\text{50 m} \over D} \sqrt{\text{1 AU} \over R}
		{\text{3 g/cm}^3 \over \rho_{ast}}\text{cm/s},
	\end{split}
\end{equation}
where the dimensionless parameter $g < 1$ accounts for the difference
in reflectivity between the coated and uncoated hemispheres, the integration
of the component of the radiation pressure force perpendicular to the radius
vector along the elliptic orbit, the projection of the spin angular momentum
direction onto the orbital plane and along the semiminor axis and the
variation of $R$ along the orbit reducing the radiation intensity away from
perihelion.

In a single perihelion passage the speed $v$ (Eq.~\ref{v}) is replaced by
$v + \Delta v$ and the energy $E$ is replaced by
\begin{equation}
	E^\prime \approx E + v \Delta v = v^2 \left({1 \over 2} - {1 \over
	1+e} + \delta\right),
\end{equation}
to lowest order in
\begin{equation}
	\delta \equiv {\Delta v \over v} \approx {g \over 0.4} {3 \times
	10^{-9} \over \sqrt{1+e}}{\text{3 g/cm}^3 \over \rho_{ast}}
	{\text{50 m} \over D}.
\end{equation}
Using Eq.~\ref{v} for $E$,
\begin{equation}
	{E^\prime \over E} \approx 1 - 2\delta {1+e \over 1-e},
\end{equation}
\begin{equation}
	{a^\prime \over a} = {E \over E^\prime} \approx 1 + 2\delta
	{1+e \over 1-e}
\end{equation}
and the mean angular motion $n$ becomes $n^\prime$:
\begin{equation}
	{n^\prime \over n} = \left({a^\prime \over a}\right)^{-3/2} \approx
	1 - 3\delta {1+e \over 1-e}.
\end{equation}
After $t/P$ orbits,
\begin{equation}
	n(t) \approx n(0)\left(1 - 3\delta{1+e \over 1-e}{t \over P}\right).
\end{equation}

The mean anomaly $M$ advances at a rate $dM/dt = n(t) = 2\pi/P(t)$ and
drifts from the unperturbed motion
\begin{equation}
	{d(M^\prime - M) \over dt} = n^\prime - n \approx 3 n \delta
	{1+e \over 1-e} {t \over P}.
\end{equation}
After a time $T$ the cumulative drift
\begin{equation}
	M^\prime - M \approx {3 \over 2}n \delta {1+e \over 1-e}
	{T^2 \over P}.
\end{equation}
Near Earth encounter at a distance $R \approx\,$1 AU this corresponds to a
displacement
\begin{equation}
	\begin{split}
	\Delta X &\sim (M^\prime - M)R \approx {3 \over 2}{1+e \over 1-e}
	n \delta {RT^2 \over P}\\ &\approx 3 \pi R {g \over 0.4} 3 \times
	10^{-9} {\sqrt{1+e} \over 1-e} \left({T \over P}\right)^2 
	{\text{3 g/cm}^3 \over \rho_{ast}} {\text{50 m} \over D}\\ 
	&\approx {g \over 0.4} 1.2 \times 10^9 \left({T \over \text{100 y}}
	\right)^2 \left({\text{3 y} \over P}\right)^2{\text{3 g/cm}^3 \over
	\rho_{ast}}{\text{50 m} \over D}\ \text{cm},
	\end{split}
\end{equation}
where the numerical result takes $e = 0.5$.  A complete miss ($\Delta X >
R_E$) occurs at
\begin{equation}
	T > 75 \sqrt{{0.4 \over g}{\rho_{ast} \over 3\text{g/cm}^3}
	{D \over \text{50 m}}}{P \over \text{3 y}}\,\text{y}.
\end{equation}
\section{Material Requirements}
Measurements \cite{N16} of the reflectivity of alkali metals describe bulk
metal.  The intensity skin depth in Li and Na is $\approx 180\,$\AA\ at the
visible and near-infrared wavelengths at which nearly all the Solar energy
is radiated \cite{M97}.  As a result, a layer $h = 0.1\,\mu$ thick has, to
better than 1\% accuracy, the same reflectivity as a half-space even at
normal incidence (and more accurately as other angles of incidence).   The
mass required to coat a sphere of diameter $D$ with metal of density
$\rho_m$
\begin{equation}
	M_{metal} = \pi D^2 \rho_m h
	\approx
	\begin{cases}
		426\left({D \over \text{50 m}}\right)^2\text{g}&\text{Li}\\
		763\left({D \over \text{50 m}}\right)^2\text{g}&\text{Na.}\\
	\end{cases}
\end{equation}

It would be necessary to deposit this material as an atomic vapor from a
spacecraft orbiting the asteroid because impact with relative velocity $>
10\,$km/s of metal droplets released in a flyby would produce
micro-fireballs, and essentially all the metal would be lost in their
expansion.  Atomic vapor would be deposited efficiently, even at this (or
higher) relative speed, but it would be difficult to produce a vapor cloud
confined so that a significant fraction of it would strike the asteroid,
unless dispensed from a platform with low relative velocity.  A platform in
polar orbit close to the asteroid could dispense vapor and coat any desired
portion of the surface.

Solar energy could be used to evaporate the metal.  If this is to occur in a
time $\Delta t \ll P$, the required power would be
\begin{equation}
	{\cal P} = {\pi D^2 \Delta H_{v} \rho_m h \over \mu \Delta t} =
	\begin{cases}
		88 \left({D \over \text{50 m}}\right)^2
		\left({10^5\,\text{s} \over \Delta t}\right) \text{W} &
		\text{Li}\\
		32 \left({D \over \text{50 m}}\right)^2
		\left({10^5\,\text{s} \over \Delta t}\right) \text{W} &
		\text{Na},
	\end{cases}
\end{equation}
where $\Delta H_{v}$ is the latent heat of evaporation (145 kJ/mole for Li
and 97 kJ/mole for Na) and $\rho_m$ is the density of the metal (0.54
g/cm$^3$ for Li and 0.97 g/cm$^3$ for Na) and $\mu$ its molecular weight.
The required collecting area would be tiny:
\begin{equation}
	A = {{\cal P} \over I_\odot \varepsilon}
	\left({R \over \text{1 AU}}\right)^2 =
	\begin{cases}
		{0.06 \over \varepsilon} \left({R \over \text{1 AU}}\right)^2
		\left({D \over \text{50 m}}\right)^2 \left({10^5\,\text{s}
		\over \Delta t}\right) \text{m}^2 & \text{Li}\\
		{0.02 \over \varepsilon} \left({R \over \text{1 AU}}\right)^2
		\left({D \over \text{50 m}}\right)^2 \left({10^5\,\text{s}
		\over \Delta t}\right) \text{m}^2 & \text{Na},\\
	\end{cases}
\end{equation}
where $\varepsilon$ is the efficiency of conversion of the Solar intensity
$I_\odot$ to latent heat of evaporation of the metal.  For a reflector
focusing Solar radiation (a solar furnace) $\varepsilon$ may approach unity,
while for Solar cells $\varepsilon \approx 0.2$ is likely.  Using Solar
energy to evaporate enough metal to coat a threatening asteroid is feasible.
\section{Discussion}
With the expected century of warning of Tunguska-class (50 m diameter)
asteroidal threats, it is feasible to change their Solar reflectivity and
radiation pressure force to move their impact to locations where it would do
little harm, or to avoid impact entirely.  This may be done by coating the
asteroid with a sub-micron layer of reflective alkali metal that would
reduce the effective force of Solar gravity on it, or provide a torque that
would systematically increase or decrease its orbital angular momentum.

In comparison to impact or an ion engine, the method is particularly
advantageous against larger (km-class) asteroids that are rarer but whose
impact would be catastrophic.  For an asteroid with diameter of 1 km a few
hundred kg of metal would suffice to provide a reflective coat.  Such larger
asteroids would take longer to deflect, but with a systematic torque the
time required to avoid impact would be proportional to the square root of
the asteroidal diameter (the sixth root of its mass), and would be $< 1000$
years for km-class asteroids.  In contrast, deflection by impact or by an
attached ion engine would require a mass of impactor or ionic propellant
proportional to that of the asteroid, and to the cube of its diameter.
\section*{Conflict of Interest Statement}
The author states that there is no conflict of interest.
\section*{Acknowledgments}
I thank D. Eardley, R. L. Garwin and K. Pister for discussions.


\begin{thebibliography}{99}
	\bibitem{T21} Tunguska event 
		\url{https://en.wikipedia.org/wiki/Tunguska_event}
		accessed Nov. 25, 2021.
	\bibitem{S83} Shoemaker, E. 1983 Asteroid and Comet Bombardment of
		the Earth, Ann. Rev. Earth Planet. Sci. 11 (1) 461--494.
	\bibitem{B02} Brown, P., Spalding, R. E., ReVelle, D. O. {\it et
		al.\/} 2002 The flux of small near-Earth objects colliding
		with the Earth, Nature 420 (6913) 294--296.
	\bibitem{B21} Brucker, M. J., Larsen, J., McMillan, R. S. {\it et
		al.\/} 2021 SPACEWATCH (R) NEO Astrometry and
		Characterization, Bull.~AAS 53(7) 101.01.
	\bibitem{A21} Abell, P.~A., Raymond, C., Daly, T. {\it et al.\/}
		2021 Near-Earth Object Characterization Priorities and
		Considerations for Planetary Defense, Planetary Science
		and Asteroidal Decadal Survey 2023--2032, Bull.~AAS 53 (4)
		e-id.~270.
	\bibitem{M21} Mainzer, A., Abell, P., Bannister, M.~T. {\it et
		al.\/} 2021 The Future of Planetary Defense in the Era of
		Advanced Surveys, Planetary Science and Asteroidal Decadal
		Survey 2023--2032, Bull.~AAS 53 (4) e-id.~259.
	\bibitem{H21} Heinze, A. N., Denneau, L., Tonry, J. L. 2020 NEO
		Population, Velocity Bias, and Impact Risk from an ATLAS
		Analysis, Planet. Sci. J. 2 (1), 12.
	\bibitem{H10a} Hyland, D. C., Altwaijry, H. A., Ge, S. {\it et
		al.\/} 2010 A Permanently-Acting NEA Damage Mitigation
		Technique via the Yarkovsky Effect, Cosmic Res. 48 (5)
		430--436.
	\bibitem{H10b} Hyland, D. C., Altwaijry, H. A., Margulieux, R. {\it
		et al.\/} 2010 A Mission Template for Exploration and Damage
		Mitigation of Potential Hazard of Near Earth Asteroids,
		Cosmic Res. 48 (5) 437--442.
	\bibitem{D21a} 65803 Didymos
		\url{https://en.wikipedia.org/wiki/65803_Didymos} accessed
		December 5, 2021.
	\bibitem{D21b} Double Asteroid Redirection Test
		\url{https://en.wikipedia.org/wiki/}
		\url{Double_Asteroid_Redirection_Test}
		accessed December 5, 2021.
	\bibitem{N16} Nathanson, J.~B. 1916 The Reflecting Power of the
		Alkali Metals, Ap.~J. XLIV (3) 137--168.
	\bibitem{M97} McWhirter, J.~D. 1997 Extinction Coefficient and Skin
		Depth of Alkali Metals from 10 to 1000 nm, Optics and Lasers
		in Engineering 28, 305--309.
\end{thebibliography}
\end{document}